\documentclass[preprint]{elsarticle}

\usepackage{lineno,hyperref}
\usepackage{amsmath}
\usepackage{amssymb}
\usepackage{graphicx}
\usepackage{esint}
\modulolinenumbers[5]

\journal{Journal of \LaTeX\ Templates}

\begin{document}

\begin{frontmatter}

\title{Non-Markovian Dynamics of  Quantum Open Systems Embedded in a Hybrid Environment}

\author{Xinyu Zhao$^{1}$}
\author{Wufu Shi$^{1}$}
\author{J. Q. You$^{2,3}$}
\author{Ting Yu$^{1,2,4}$\corref{cor1}}
\address{$^{1}$Department of Physics
and Engineering Physics, Stevens Institute of Technology, Hoboken,
New Jersey 07030, USA}
\address{$^{2}$Beijing Computational Science Research Center, Beijing 100094, China}
\address{$^{3}$Synergetic Innovation Center of Quantum Information and Quantum Physics, University of Science and Technology of China, Hefei, Anhui 230026, China}
\address{$^{4}$School of Physics and Optoelectronic Engineering, Yangtze University, Jingzhou 434023, China}


\cortext[cor1]{Corresponding author's email:  Ting.Yu@stevens.edu}


\begin{abstract}
Quantum systems of interest are typically coupled to several quantum channels (more generally  environments). 
 In this paper, we develop an exact stochastic Schr\"{o}dinger equation 
for an open  quantum system coupled to a hybrid environment containing both bosonic and fermionic 
particles.  Such a stochastic differential equation may be obtained directly from a microscopic model
through  employing  a classical complex Gaussian noise  and a non-commutative fermionic noise to simulate
the hybrid bath. 
As an immediate application of our developed stochastic approach, we show that the evolution of the reduced
density matrix can be derived by taking the average over both the bosonic noise and the fermionic noise.
Three specific examples are given in this paper to illustrate that the hybrid quantum trajectory 
is fully consistent with the standard quantum mechanics. Our examples also shed new light on the special features 
exhibited by the fermionic bath and bosnoic bath.
\end{abstract}


\end{frontmatter}

\nolinenumbers
\section{Introduction}

A quantum system, when it is not isolated,  can be in contact
with several types of environments.  Physically, such open quantum systems  like an electron relaxation in
a solid may interact with a bosonic system and be coupled to
some fermionic systems at the same time \cite{hybrid1,hybrid2,hybrid3}.  In a similar manner, one can recognize 
that an atomic system of interest can be coupled to both classical laser fields and quantized radiation fields \cite{Gardiner1}.  
Therefore, a hybrid quantum 
open system theory is potentially useful since it provides a systematic 
approach to dealing with the dynamics of an open system coupled to multiple environments in a direct manner.
Fundamentally, the dynamics of open quantum systems
embedded in one or more environments has attracted
the wide-spread interest in recent years \cite{Unrhu,Breuer,Xinyu2011,Nlevel}. On the one hand,
the temporal behaviours of quantum open systems are essential for understanding many fundamental
issues of quantum theory such as quantum dissipation and decoherence \cite{Deco1,Deco2,Deco3,Deco4,Yu-Eberly04,Nqubit,Ncavity}.
On the other hand, many novel applications based on quantum devices
also require a better understanding on the interaction between the quantum
system of interest and its environment in order to manipulate and control the system's dynamics \cite{DD,FBC}.
Although a realistic environment can be very complicated, it is typically composed of bosons and fermions. For a bosonic bath, a set
of powerful tools have been developed to investigate the open system
dynamics, such as path integral approach \cite{Feynman-Vernon,ZhangWM-2cav},
master equation approach \cite{H-P-Z,Hu1,Hu2,Leggett}, and Markov and non-Markovian
quantum trajectory approach \cite{Gisin-Percival,Dalibardetal,QSD,Yu1999,YuQBM}. For
fermionic bath, similar tools have also been developed, including
scattering theory \cite{scattering}, non-equilibrium Green's function
approach \cite{NEGF}, and fermionic path integral \cite{ZhangDQD,Zhang2012PRL}. 
Notably, the fermionic quantum state diffusion equations have been developed recently 
\cite{ZhaoFB,ShiFB,ChenFB}.  Although bosonic and fermionic formalisms  share many
similarities, a unified description of both types
of baths is still useful for the purpose of direct applications.

\begin{figure}
\begin{centering}
\includegraphics[width=1\columnwidth]{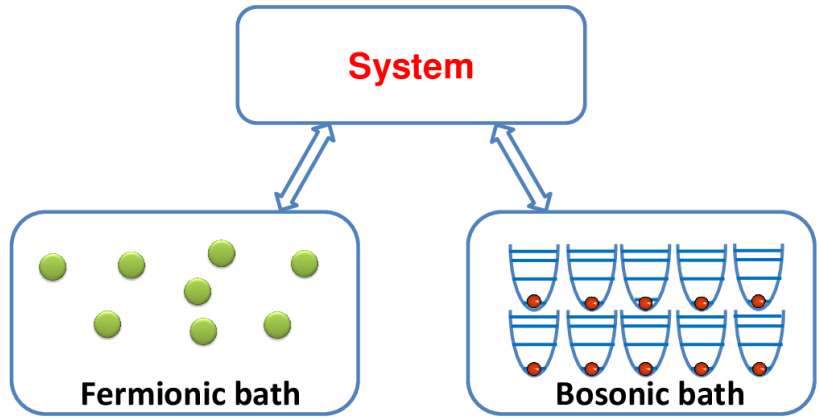}
\end{centering}

\caption{(Color online) Schematic diagram of a quantum system coupled to a fermionic bath and a bosonic bath simultaneously.}

\label{Fig1}
\end{figure}

In this paper, we will consider a hybrid case that the environment
is composed of both bosons and fermions as shown in Fig. \ref{Fig1}. 
Of many applications is the  primary example  of quantum dot model where the quantum dots may interact
with two fermionic reservoirs (source and drain) and other agents such as 
a phonon bath. In this context, the dynamics of the quantum dot system is
determined by both the fermionic reservoirs and the bosonic bath. 
The theoretical approach to be developed in this paper will be
capable of taking into account of  the environmental effects arising from both
types of environments.  We shall show that the non-Markovian quantum state
diffusion (NMQSD) approach is applicable to this extended case.
In this method, the dynamic evolution
of the open system is decomposed into an ensemble of quantum trajectories of pure
states, and the reduced density matrix is described by the statistical
average over these generated trajectories. It is worth noting the fact that
this method is well developed for both bosonic bath and fermionic
bath, and the method becomes one of the most hopeful candidates
to solve the hybrid bath problem. For the case of bosonic bath, several interesting
physical systems have been studied in the past fifteen years \cite{QSD,Yu1999,YuQBM,Yu-FiniteT,Jing-Yu2010}.
As a fundamentally theoretical study, it has been developed from solving
single-particle system to solving many-body systems \cite{Nlevel,Nqubit,Ncavity}.
Furthermore, as a computing tool in real applications \cite{QSDHierachy,Lam},
the NMQSD approach also showed its potential value in many interesting
problems including precision quantum measurement \cite{Chen}, quantum control
dynamics \cite{JingPQ}, and quantum biology \cite{Eisfeld2011}.
In either the bosonic or fermionic NMQSD approach,
the central idea is to encode  all the influences  of the environments
on the system into a set of classical random variables forming a stochastic 
process. Taking the statistic average over the stochastic variables
is equivalent to taking the partial trace over the environment to
obtain the reduced density matrix. The difference between bosonic and fermionic approaches 
is that a bosonic bath can be represented by a complex Gaussian noise,
while the fermionic bath is simulated by a non-commutative Grassmann noise. 
For a hybrid bath, an exact dynamical equation will typically contain both the complex Gaussian
noise and Grassmann Gaussian noise.
 
The paper is organized as follows. In Sec.~\ref{sec:II}, we describe
a model of hybrid bath and point out some new features arising from 
the hybrid bath case where the fermionic bath is assumed to commutes or anti-commutes
with the system of interest. In Sec.~\ref{sec:III}, we analyze the commutative
case with two examples. First, a general NMQSD equation and the corresponding
master equation are derived. Then, we examine our general formalism by exactly solving
a simple example of  the single qubit dissipative model. It is shown
that, as expected,  the result predicted by the newly developed NMQSD approach is
identical to the solution based on the ordinary quantum mechanics.
Moreover, the two-qubit dissipative model is also studied for the hybrid bath case.
 Sec.~\ref{sec:IV} is devoted to investigating the anti-commutative case. Again, a general NMQSD approach 
 can be developed.  As an important example,  we show how to use the new approach to study 
 the Anderson model in the hybrid bath context. We identify the different
impacts of fermionic bath and bosonic bath on the system dynamics. We conclude in
Sec.~\ref{sec:V}.

\section{\label{sec:II} Two Types of Hybrid Bath: Commutative and Anti-commutative}

An open system embedded in a hybrid bath may be described by the following
Hamiltonian
\begin{equation}
H_{tot}=H_{S}+H_{FB}+H_{BB}+H_{FI}+H_{BI},\label{Hybrid}
\end{equation}
where $H_{S}$ describes the Hamiltonian of the system,
\begin{equation}
H_{BB}=\sum_{r}\Omega_{r}b_{r}^{\dagger}b_{r},\;
 H_{FB}=\sum_{k}\epsilon_{k}c_{k}^{\dagger}c_{k},\label{HB}
\end{equation}
are the bosonic bath and the fermionic bath respectively, where ``$b_{r}$'' and ``$c_{k}$''
are the annihilation operators for a single mode of bosonic bath and
fermionic bath respectively. The interaction between system and two
baths is given by
\begin{equation}
H_{BI}=\sum_{r}\lambda_{r}b_{r}^{\dagger}L_{b}+{\rm H.c.},\;
H_{FI}=\sum_{k}\mu_{k}c_{k}^{\dagger}L_{f}+{\rm H.c.},\label{HInt}
\end{equation}
where $L_{b}$ and $L_{f}$ are the bosonic and fermionic coupling operators.
Typically, the bosonic bath commutes with both the fermionic bath
and the system no mater the system is composed of fermions or bosons.
However, the commutation relation between the system and the fermionic
bath could fall into two different categories. Depending
on the commutation relation between the system and the fermionic bath
(commutative or anti-commutative), the physical model for this Hamiltonian
and the technique of solving this model are totally different. Therefore,
we need to develop two parallel schemes to deal with these two different
cases.


\subsection*{Case 1: System commutes with fermionic bath.}

The Hamiltonian (\ref{Hybrid})  for this case typically describes
an effective fermionic bath. For example, a spin-chain
bath can be transformed into an effective fermionic bath by using the
Jordan-Wigner transformation and the Fourier transform \cite{ZhaoFB,Barouch}. After the transformation, t
hese effective fermions satisfying fermionic commutation relations. However, since the original spins living in a Hilbert 
space separated from the system degree of freedom, they all commute with the system.
Therefore, in the case of effective fermionic bath, the creation and annihilation
operators $c_{k}^{\dagger}$ and $c_{k}$ will commute with any operators
living in the Hilbert space of the system $H_{S}$. 

\subsection*{Case 2: System anti-commutes with fermionic bath.}

The anti-commutative case naturally arises when both the system and  the bath  are composed 
of a set of electrons.  A well-known example is a quantum dot connected to a source and a drain reservoirs, where 
the system Hamiltonian of this model is $H_{S}=\omega_{d}d^{\dagger}d$. Obviously, the annihilation
operator ``$d$'' for the system and the operators ``$c_{k}$'' for
the bath satisfy the anti-commutation relation $\{d,c_{k}\}=0$ and $\{d^\dagger,c_{k}=0\}$. 

We will develop two different schemes in the following sections  for the two cases described above. Several specific examples
are provided.

\section{\label{sec:III} Commutative Case}

\subsection{\label{sub:IIIA}General Stochastic Schr\"odinger Equation}

First, we consider the commutative case.   In this case, the total Hamiltonian can be transformed
into the interaction picture as
\begin{equation}
H_{tot}^{int}=H_{S}+(\sum_{k}\mu_{k}c_{k}^{\dagger}L_{f}e^{i\epsilon_{k}t}+\sum_{r}\lambda_{r}b_{r}^{\dagger}L_{b}e^{i\Omega_{r}t}+\mathrm{H.c.}).
\end{equation}
By introducing multi-mode bosonic coherent states and fermionic coherent
states
\begin{equation}
|z\rangle=\prod_{r}\exp\left\{ z_{r}b_{r}^{\dagger}\right\} |0\rangle,
\end{equation}
\begin{equation}
|\xi\rangle=\prod_{k}\left(1-\xi_{k}c_{k}^{\dagger}\right)|0\rangle,
\end{equation}
the stochastic state vector can be defined as
\begin{equation}
|\psi_{t}(z^{*},\xi^{*})\rangle=\langle z^{*},\xi^{*}|\psi_{tot}(t)\rangle.
\end{equation}
Throughout the paper, we will use the short-notation $|\psi_{t}\rangle\equiv|\psi_{t}(z^{*},\xi^{*})\rangle$
if no confusion arises. Because of the different properties of bosonic
coherent states and fermionic coherent states \cite{ZhangRMP}, the
noise variables introduced here are rather different. For bosonic coherent
states, $z_{r}$ is an ordinary complex variable, while for fermionic coherent
states, $\xi_{k}$ is a Grassmann variable satisfying anti-commutative
relations $\{\xi_{i},\xi_{j}\}=0$. Starting with the Schr\"{o}dinger
equation for the total system,  one can derive the dynamic equation for the stochastic
state vector,
\begin{eqnarray}
\frac{\partial}{\partial t}|\psi_{t}\rangle & = & -i\langle z^{*},\xi^{*}|H_{tot}^{int}(t)|\psi_{tot}(t)\rangle\nonumber \\
 & = & [-iH_{S}+L_{f}\xi_{t}^{\ast}-L_{f}^{\dagger}\int dsK_{f}(t,s)\frac{\delta_{l}}{\delta\xi_{s}^{\ast}}\nonumber \\
 &  & +L_{b}z_{t}^{\ast}-L_{b}^{\dagger}\int dsK_{b}(t,s)\frac{\delta}{\delta z_{s}^{\ast}}]|\psi_{t}\rangle,\label{QSDC}
\end{eqnarray}
where $K_{b}(t,s)=\sum_{r}\lambda_{r}^{2}e^{-i\Omega_{r}(t-s)}$ and
$K_{f}(t,s)=\sum_{k}\mu_{k}^{2}e^{-i\epsilon_{k}(t-s)}$ are the correlation
functions for the bosonic bath and  the fermionic bath,  respectively. The equation
(\ref{QSDC}) is the fundamental equation governing the dynamics
of the stochastic state vector $|\psi_{t}\rangle$. Note that this equation contains two types of noises as
\begin{equation}
z_{t}^{*}=-i\sum_{r}z_{r}^{*}e^{i\Omega_{r}t},
\end{equation}
\begin{equation}
\xi_{t}^{*}=-i\sum_{k}\xi_{k}^{*}e^{i\epsilon_{k}t},
\end{equation}
where $z_{t}^{*}$ is a complex Gaussian process and $\xi_{t}^{*}$
is a Grassmann stochastic process. They satisfy the following statistical relations
\begin{equation}
\langle z_{t}\rangle_{b}=\langle z_{t}^{*}\rangle_{b}=0,\;\langle z_{t}z_{s}^{*}\rangle_{b}=K_{b}(t,s),
\end{equation}
\begin{equation}
\langle\xi_{t}\rangle_{f}=\langle\xi_{t}^{*}\rangle_{f}=0,\;\langle\xi_{t}\xi_{s}^{*}\rangle_{f}=K_{f}(t,s).
\end{equation}
The statistical averages over  both the complex noise and Grassmann noises  are
defined as $\langle\cdot\rangle_{b}=\int\prod_{r}\frac{1}{\pi}e^{-|z_{r}|^{2}}dz_{r}^{2}[\cdot]$
and $\langle\cdot\rangle_{f}=\int\prod_{k}d\xi_{k}^{*}d\xi_{k}e^{-\xi_{k}^{*}\xi_{k}}[\cdot]$,
respectively.

In order to solve  Eq.~(\ref{QSDC}), one has to deal with the functional
derivatives. Similar to the technique used in Refs.~\cite{Yu1999,ShiFB,ZhaoFB,ChenFB},
we can always replace these functional derivatives by some time-dependent
operators $O$ and $Q$ as
\begin{equation}
\frac{\delta}{\delta z_{s}^{*}}|\psi_{t}\rangle=O(t,s,z^{*},\xi^{*})|\psi_{t}\rangle,
\end{equation}
\begin{equation}
\frac{\delta}{\delta\xi_{s}^{*}}|\psi_{t}\rangle=Q(t,s,\xi^{*},z^{*})|\psi_{t}\rangle.
\end{equation}
Then, the NMQSD equation (\ref{QSDC}) can be written in a more compact
form,
\begin{equation}
\frac{\partial}{\partial t}|\psi_{t}\rangle=[-iH_{s}+L_{f}\xi_{t}^{\ast}-L_{f}^{\dagger}\bar{Q}+L_{b}z_{t}^{*}-L_{b}^{\dagger}\bar{O}]|\psi_{t}\rangle,\label{QSDC2}
\end{equation}
where $\bar{O}(t,z^{*},\xi^{*})=\int_{0}^{t}K_{b}(t,s)O(t,s,z^{*},\xi^{*})ds$,
$\bar{Q}(t,z^{*},\xi^{*})=\int_{0}^{t}K_{f}(t,s)Q(t,s,z^{*},\xi^{*})ds$.
If these time-dependent operators $O$ and $Q$ can be determined,
the NMQSD equation will take a time-local form and the equation can be solved
numerically in a more straightforward way. The consistency 
condition may be employed,
\begin{equation}
\frac{\partial}{\partial t}\frac{\delta}{\delta z_{s}^{*}}|\psi_{t}\rangle=\frac{\delta}{\delta z_{s}^{*}}\frac{\partial}{\partial t}|\psi_{t}\rangle,
\end{equation}
\begin{equation}
\frac{\partial}{\partial t}\frac{\delta}{\delta\xi_{s}^{*}}|\psi_{t}\rangle=\frac{\delta}{\delta\xi_{s}^{*}}\frac{\partial}{\partial t}|\psi_{t}\rangle,
\end{equation}
and gives rise to
\begin{eqnarray}
\frac{\partial}{\partial t}O & = & [-iH_{s}+L_{f}\xi_{t}^{\ast}-L_{f}^{\dagger}\bar{Q}+L_{b}z_{t}^{*}-L_{b}^{\dagger}\bar{O},O]\nonumber \\
 &  & -L_{b}^{\dagger}\frac{\delta}{\delta z_{s}^{*}}\bar{O}-L_{f}^{\dagger}\frac{\delta}{\delta z_{s}^{*}}\bar{Q}, \label{EqO}
\end{eqnarray}

\noindent 
\begin{eqnarray}
\frac{\partial}{\partial t}Q & = & [-iH_{s},Q]-\{L_{f}\xi_{t}^{*},Q\}+[L_{b}z_{t}^{*},Q]\nonumber \\
 &  & -L_{f}^{\dagger}\bar{Q}(-\xi^{*})Q+QL_{f}^{\dagger}\bar{Q}-L_{b}^{\dagger}\bar{O}(-\xi^{*})Q\nonumber \\
 &  & +QL_{b}^{\dagger}\bar{O}-L_{b}^{\dagger}\frac{\delta_{l}}{\delta\xi_{s}^{\ast}}\bar{O}-L_{f}^{\dagger}\frac{\delta_{l}}{\delta\xi_{s}^{*}}\bar{Q}, \label{EqQ}
\end{eqnarray}
\noindent with the initial conditions 
\begin{equation}
O(t,t,z^{*},\xi^{*})=L_{b},
\end{equation}
\begin{equation}
Q(t,t,z^{*}\xi^{*})=L_{f}.
\end{equation}
Given these conditions, the $O$ and $Q$ operators can be fully determined,
as a result, the NMQSD equation (\ref{QSDC2}) becomes more useful for the analytical  
purpose and numerical simulations. However,
A single solution of the NMQSD equation can not fully describe the dynamic
evolution of the system. Actually, it only gives one possible realization
of many possible solutions of  the NMQSD equation corresponding a specific sample
path taken by the stochastic process  $z_{t}^{*}$
and $\xi_{t}^{*}$. In order to obtain the full picture of the evolution
of the system, we need to reproduce the reduced density matrix from
the stochastic state vector $|\psi_{t}\rangle$ as
\begin{equation}
\rho(t)=\langle\langle P_{t}\rangle_{f}\rangle_{b},\label{ReproduceRho}
\end{equation}
where $P_{t}\equiv|\psi_{t}(z^{*},\xi^{*})\rangle\langle\psi_{t}(z^{*},-\xi^{*})|$
is the stochastic density operator. Given the relation Eq. (\ref{ReproduceRho}),
the physical meaning of the NMQSD equation becomes clear. By choosing
a random realization of the noises $z_{t}^{*}$ and $\xi_{t}^{*}$
(reflecting the states of the environment), the evolution of the reduced
density matrix is decomposed into many pure-state quantum trajectories
$|\psi_{t}\rangle$. However, taking the statistical average over
all of these trajectories, the reduced density matrix is reproduced.
Therefore, the complicated properties of the environment are all encoded
into noise functions $z_{t}^{*}$ and $\xi_{t}^{*}$, so that tracing
out the environment is equivalent to taking average over all the realizations
of the noises. Based on the relation (\ref{ReproduceRho}), the master
equation can be derived as
\begin{eqnarray}
\frac{d}{dt}\rho & = & -i[H_{S},\rho]+[L_{b},\langle\langle P_{t}\bar{O}^{\dagger}\rangle_{f}\rangle_{b}]+[\langle\langle\bar{O}P_{t}\rangle_{f}\rangle_{b},L_{b}^{\dagger}]\nonumber \\
 &  & +[L_{f},\langle\langle P_{t}\bar{Q}^{\dagger}(-\xi)\rangle_{f}\rangle_{b}]+[\langle\langle\bar{Q}P_{t}\rangle_{f}\rangle_{b},L_{f}^{\dagger}], \label{MEQ1}
\end{eqnarray}
where the Novikov theorem for fermionic case \cite{ZhaoFB} and bosonic
case \cite{Yu1999} are used in the derivation. Although taking the
statistical averages $\langle\cdot\rangle_{b}$ and $\langle\cdot\rangle_{f}$
are not simple in the general case, there is still a special case.
When the operators $O$ and $Q$ are noise-independent, the master
equation can be written in a simpler form as
\begin{equation}
\frac{d}{dt}\rho=-i[H_{S},\rho]+\{[\bar{Q}\rho,L_{f}^{\dagger}]+[\bar{O}\rho,L_{b}^{\dagger}]+\mathrm{H.c.}\},\label{eq:MEQ2}
\end{equation}
Actually, this special case is very common in many interesting models
\cite{QSD,ZhaoFB,Yu1999,Ncavity} in which the exact $O$ or $Q$
operators just contain no noises. Moreover, in general case, we can
still expand $O$ and $Q$ into functional series and only taking
the first term (with zeroth order of noise variables) of the expansions
as $O(t,s,z^{*},\xi^{*})\approx O^{(0)}(t,s)$ and $Q(t,s,z^{*},\xi^{*})\approx Q^{(0)}(t,s)$.
This approximation is called the zeroth order approximation \cite{Yu1999}.
The validity and accuracy of this approximation is analyzed in Ref.
\cite{Xu2014}. Actually, the accuracy is proved to be much better
than the weak coupling approximation.

\subsection{Example 1: Two qubits in a hybrid bath}

\begin{figure}
\begin{centering}
\includegraphics[width=1\columnwidth]{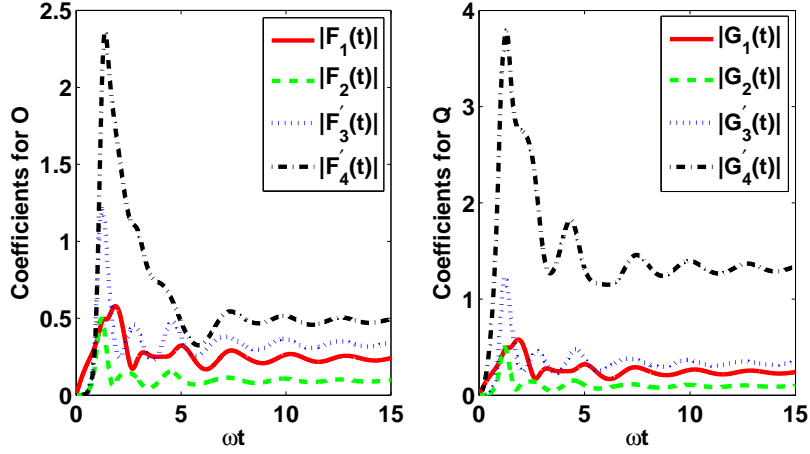}
\par\end{centering}

\caption{\label{Coeff}(Color online) Time evolution for the coefficients in
$O$ and $Q$ operators. In the left subplot, the red (solid), green
(dashed), blue (dash-dotted) and black (dash-dotted) curves are the
absolute values of the coefficients $|F_{1}(t)|$, $|F_{2}(t)|$,
$|F_{3}^{\prime}(t)|$, and $|F_{4}^{\prime}(t)|$, respectively.
In the right subplot, those curves represent $|G_{1}(t)|$, $|G_{2}(t)|$,
$|G_{3}^{\prime}(t)|$, and $|G_{4}^{\prime}(t)|$, respectively.}
\end{figure}

In order to illustrate the NMQSD approach for a hybrid bath we discussed
above, we will solve some particular examples in details. In the first
example, we will consider a two-qubit system interacting with two dissipative
baths, one is bosonic and the other fermionic. From this example, we show that dynamic equation
 for a hybrid bath is not the simple combination of a fermionic bath and 
 a bosonic bath. The cross-terms in $O$ and $Q$ operators reflect the correlation between two baths
 through interaction with the system of interest. In the
general model described by Eqs.~(\ref{Hybrid}-\ref{HInt}),
the two qubits example is the special case that 
\begin{equation}
H_{S}=\frac{\omega}{2}(\sigma_{z}^{A}+\sigma_{z}^{B}),
\end{equation}
\begin{equation}
L_{b}=L_{f}=\sigma_{-}^{A}+\kappa_{B}\sigma_{-}^{B}.\label{L2qu}
\end{equation}
where $\kappa_{B}$ is a parameter describing the coupling strength
between the second qubit and the baths. We will first investigate
the case with $\kappa_{B}=1$ in this subsection.  A special case with $\kappa_{B}=0$ will be considered
later, which means the second qubit evolves independently from the
other part of the total system and the model reduces to a single qubit case. Given this
specific model, the NMQSD equation can be formally written as
\begin{eqnarray}
\frac{\partial}{\partial t}|\psi_{t}\rangle & = & [-i\frac{\omega}{2}(\sigma_{z}^{A}+\sigma_{z}^{B})+(\sigma_{-}^{A}+\sigma_{-}^{B})(\xi_{t}^{\ast}+z_{t}^{*})\nonumber \\
 &  & -(\sigma_{+}^{A}+\sigma_{+}^{B})(\bar{Q}+\bar{O})]|\psi_{t}\rangle. \label{QSD2Q}
\end{eqnarray}
In fact, it is instructive to compare this dynamic equation for a hybrid bath with the model that two qubits interact with either a single bosonic bath or a single fermionic bath. For a single bath, the NMQSD equation should be
$\frac{\partial}{\partial t}|\psi_t\rangle=[-iH_S+L_b z_t^\ast-L_b^\dagger\bar{O}]|\psi_t\rangle$ (bosonic \cite{Xinyu2011}) or
$\frac{\partial}{\partial t}|\psi_t\rangle=[-iH_S+L_f \xi_t^\ast-L_f^\dagger\bar{Q}]|\psi_t\rangle$ (fermionic \cite{ZhaoFB}). It seems that Eq.~(\ref{QSD2Q}) is nothing more than a direct summation of a fermionic bath and a bosonic bath. However, more information is encoded in the $O$ and $Q$ operators. According to Eqs. (\ref{EqO}-\ref{EqQ}), the exact $O$ and $Q$ operators can be determined as
\begin{eqnarray}
O & = & f_{1}(t,s)O_{1}+f_{2}(t,s)O_{2}+i\int_{0}^{t}ds^{\prime}f_{3}(t,s,s^{\prime})z_{s^{\prime}}^{*}O_{3}\nonumber \\
 &  & +i\int_{0}^{t}ds^{\prime}f_{4}(t,s,s^{\prime})\xi_{s^{\prime}}^{*}O_{4},
\end{eqnarray}
\begin{eqnarray}
Q & = & g_{1}(t,s)Q_{1}+g_{2}(t,s)Q_{2}+i\int_{0}^{t}ds^{\prime}g_{3}(t,s,s^{\prime})z_{s^{\prime}}^{*}Q_{3}\nonumber \\
 &  & +i\int_{0}^{t}ds^{\prime}g_{4}(t,s,s^{\prime})\xi_{s^{\prime}}^{*}Q_{4}.
\end{eqnarray}
The basis operators are $O_{1}=\sigma_{-}^{A}+\sigma_{-}^{B}$, $O_{2}=(\sigma_{z}^{A}+\sigma_{z}^{B})(\sigma_{-}^{A}+\sigma_{-}^{B})$,
$O_{3}=O_{4}=\sigma_{-}^{A}\sigma_{-}^{B}$, $Q_{1}=\sigma_{-}^{A}+\sigma_{-}^{B}$,
$Q_{2}=(\sigma_{z}^{A}+\sigma_{z}^{B})(\sigma_{-}^{A}+\sigma_{-}^{B})$,
$Q_{3}=Q_{4}=\sigma_{-}^{A}\sigma_{-}^{B}$. The time-dependent coefficients
satisfy the following relations
\begin{equation}
\frac{\partial}{\partial t}f_{1}(t,s)=i\omega f_{1}+4f_{1}F_{2}+4f_{1}G_{2}+iF_{3}+iG_{3},
\end{equation}
\begin{eqnarray}
\frac{\partial}{\partial t}f_{2}(t,s) & = & i\omega f_{2}+f_{1}(4F_{2}+4G_{2}-F_{1}-G_{1})-\frac{i}{2}F_{3}\nonumber \\
 &  & +f_{2}(2F_{1}+2G_{1}-4F_{2}-4G_{2})-\frac{i}{2}G_{3},
\end{eqnarray}
\begin{eqnarray}
\frac{\partial}{\partial t}f_{3}(t,s,s^{\prime}) & = & 2i\omega f_{3}+2f_{1}F_{3}+2f_{1}G_{3}-4f_{2}F_{3}\nonumber \\
 &  & -4f_{2}G_{3}+2f_{3}F_{1}+2f_{3}G_{1},
\end{eqnarray}
\begin{eqnarray}
\frac{\partial}{\partial t}f_{4}(t,s,s^{\prime}) & = & 2i\omega f_{4}+2f_{1}F_{4}+2f_{1}G_{4}-4f_{2}F_{4}\nonumber \\
 &  & -4f_{2}G_{4}+2f_{4}F_{1}+2f_{4}G_{1},
\end{eqnarray}
\begin{equation}
\frac{\partial}{\partial t}g_{1}(t,s)=i\omega g_{1}+4g_{1}F_{2}+4g_{1}G_{2}+iF_{3}+iG_{3},
\end{equation}
\begin{eqnarray}
\frac{\partial}{\partial t}g_{2}(t,s) & = & i\omega g_{2}+g_{1}(4F_{2}+4G_{2}-F_{1}-G_{1})-\frac{i}{2}F_{3}\nonumber \\
 &  & +g_{2}(2F_{1}+2G_{1}-4F_{2}-4G_{2})-\frac{i}{2}G_{3},
\end{eqnarray}
\begin{eqnarray}
\frac{\partial}{\partial t}g_{3}(t,s,s^{\prime}) & = & 2i\omega g_{3}+2g_{1}F_{3}+2g_{1}G_{3}-4g_{2}F_{3}\nonumber \\
 &  & -4g_{2}G_{3}+2g_{3}F_{1}+2g_{3}G_{1},
\end{eqnarray}
\begin{eqnarray}
\frac{\partial}{\partial t}g_{4}(t,s,s^{\prime}) & = & 2i\omega g_{4}+2g_{1}F_{4}+2g_{1}G_{4}-4g_{2}F_{4}\nonumber \\
 &  & -4g_{2}G_{4}+2g_{4}F_{1}+2g_{4}G_{1},
\end{eqnarray}
with the initial conditions
\begin{equation}
f_{1}(t,t)=g_{1}(t,t)=1,\quad f_{2}(t,t)=g_{2}(t,t)=0,
\end{equation}
\begin{equation}
f_{3}(t,t,s^{\prime})=f_{4}(t,t,s^{\prime})=g_{3}(t,t,s^{\prime})=g_{4}(t,t,s^{\prime})=0,
\end{equation}
\begin{equation}
g_{3}(t,s,t)=-4ig_{2}(t,s),\; g_{4}(t,s,t)=-4ig_{1}(t,s)+4ig_{2}(t,s),
\end{equation}
\begin{equation}
f_{3}(t,s,t)=f_{4}(t,s,t)=-4if_{2}(t,s).
\end{equation}
where
\begin{equation}
F_{i}(t)=\int_{0}^{t}K_{b}(t,s)f_{i}(t,s)ds,\quad(i=1,2)
\end{equation}
\begin{equation}
G_{i}(t)=\int_{0}^{t}K_{f}(t,s)g_{i}(t,s)ds,\quad(i=1,2)
\end{equation}
\begin{equation}
F_{i}(t,s^{\prime})=\int_{0}^{t}K_{b}(t,s)f_{i}(t,s,s^{\prime})ds,\quad(i=3,4)
\end{equation}
\begin{equation}
G_{i}(t,s^{\prime})=\int_{0}^{t}K_{f}(t,s)g_{i}(t,s,s^{\prime})ds,\quad(i=3,4)
\end{equation}
In this example, both $O$ and $Q$ operators contain the fermionic
noise $\xi^{*}$, but the fermionic $Q$ operator also contains
the bosonic noise $z^{*}$. Although the NMQSD equation (\ref{QSD2Q})
seems to be a direct summation of two individual baths, the $O$ and $Q$ operators
in a hybrid bath are not a simple combination of these operators obtained in
the single bath case (the exact $O$ or $Q$ operators for a single
bosonic bath or a fermionic bath can be found in Ref. \cite{Xinyu2011}
and Ref. \cite{ZhaoFB} respectively). Instead, there are many cross
terms and they are coupled to each other reflecting
the fact that \textit{the effect of a hybrid bath cannot be simply treated
as the direct summation of a fermionic bath plus a bosonic bath.}
Through the system, two baths are also coupled indirectly. Such kind
of indirect coupling can be also considered as interference between
two independent baths which has been recently discussed in Ref.~\cite{Lukin}.
With the exact solution, it is possible to investigate the interference
between fermionic bath and bosonic bath in the future research. Here, we just
show one numerical result as an example to demonstrate that the interference
between two baths can be dominant under certain conditions. In Fig.~\ref{Coeff}, we compare the time evolutions
of the coefficients in $O$ and $Q$ operators. The functions $F_{i}^{\prime}(t)$
($i=3,4$) are defined as $F_{i}^{\prime}(t)=\int ds^{\prime}K_{b}(t,s^{\prime})F_{i}(t,s^{\prime})$
($i=3,4$). Similarly, the functions $G_{i}^{\prime}(t)$ ($i=3,4$)
are defined as $G_{i}^{\prime}(t)=\int ds^{\prime}K_{f}(t,s^{\prime})G_{i}(t,s^{\prime})$
($i=3,4$). In the single bath case \cite{Xinyu2011}, $O$ operator
should not contain fermionic noise term $i\int_{0}^{t}ds^{\prime}f_{4}(t,s,s^{\prime})\xi_{s^{\prime}}^{*}O_{4}$.
However, according to the numerical results for hybrid bath, the coefficient
for fermionic noise term, $F_{4}^{\prime}(t)$ can be dominant under
certain conditions. Similarly, in the $Q$ operator, the bosonic noise
term $G_{3}^{\prime}(t)$ is also larger than $G_{1}(t)$ and $G_{2}(t)$.
These results imply that the interference between two baths can be
rather complicated and important. It is also worth noting that the results in Fig.~\ref{Coeff}
is obtained in a strongly non-Markovian environment. In a Markov case, $F_1(t)$ and $G_1(t)$ will be dominant. Thus, our exact treatment of the non-Markovian hybrid bath problem could be a valuable tool to study those properties in the future.

\noindent 

\subsection{Single qubit case: Consistency with ordinary quantum mechanics}

Although the $O$ and $Q$ operators in the two-qubit example are
rather complicated, it is still possible to find a simple form when
a special case- the single qubit case is considered, namely $\kappa_{B}=0$.
In this case, the second qubit evolves independently from all the
other parts, so that it can be removed in the interaction picture.
Therefore, the NMQSD equation is reduced to
\begin{equation}
\frac{\partial}{\partial t}|\psi_{t}\rangle=[-i\frac{\omega}{2}\sigma_{z}^{A}+\sigma_{-}^{A}(\xi_{t}^{\ast}+z_{t}^{*})-\sigma_{+}^{A}(\bar{Q}+\bar{O})]|\psi_{t}\rangle,
\end{equation}
where the exact $O$ and $Q$ operators can be determined as
\begin{equation}
O=Q=f(t,s)\sigma_{-}^{A}.
\end{equation}
The time-dependent coefficient $f(t,s)$ satisfies the equation
\begin{equation}
\frac{\partial}{\partial t}f(t,s)=[i\omega+F(t)]f(t,s),
\end{equation}
where $F(t)=\int_{0}^{t}[K_{b}(t,s)+K_{f}(t,s)]f(t,s)ds$. Finally,
the exact master equation for this model is derived as
\begin{equation}
\frac{d}{dt}\rho=-i[H_{S},\rho]+\{F(t)[\sigma_{-}\rho,\sigma_{+}]+\mathrm{H.c.}\}.\label{MEQ1qu}
\end{equation}
In general, the correlation functions $K_{b}(t,s)$ and $K_{f}(t,s)$
can be very complicated. However, here, we will use a special case to
show that the result derived from NMQSD approach is consistent with
the ordinary quantum mechanics. Consider the special case that there
are only one boson and one fermion in the bosonic bath and fermionic
bath respectively, i.e., $H_{FB}=\epsilon c^{\dagger}c$, $H_{BB}=\Omega b^{\dagger}b$.
Therefore, the correlation functions is reduced to
\begin{equation}
K_{b}(t,s)=\lambda^{2}e^{-i\Omega(t-s)},
\end{equation}
\begin{equation}
K_{f}(t,s)=\mu^{2}e^{-i\epsilon(t-s)}.
\end{equation}
In the resonance case, $\omega=\epsilon=\Omega$, $\lambda=\mu$,
the equation for $F(t)$ is
\begin{equation}
\frac{\partial}{\partial t}F(t)=2\lambda^{2}+F^{2}(t).
\end{equation}
The solution is
\begin{equation}
F(t)=\sqrt{2}\lambda\tan(\sqrt{2}\lambda t).
\end{equation}
From the master equation (\ref{MEQ1qu}), the evolution of the off-diagonal
element $\rho_{21}$ is
\begin{equation}
\frac{d}{dt}\rho_{21}(t)=i\omega\rho_{21}-F^{*}(t)\rho_{21}.
\end{equation}
Finally, the solution of $\rho_{21}(t)$ is 
\begin{equation}
\rho_{21}(t)=\rho_{21}(0)e^{i\omega t}\cos(\sqrt{2}\lambda t).\label{eq:rho21}
\end{equation}
On the other hand, it is also straightforward to solve the whole system (system plus two ``baths'') with the standard Schr\"{o}dinger equation since 
the total system 
only contains three particles. It is easy to confirm that
solving the whole system gives the identical result as we
obtained in Eq.~(\ref{eq:rho21}) by using the NMQSD approach. It confirms that our NMQSD
approach is indeed consistent with ordinary quantum mechanics as expected.

\subsection{Example 2: Single qubit with dephasing bosonic bath and dissipative
fermionic bath}

In the second example, we will investigate the case that the system
is coupled to the bosonic bath and fermionic bath in two different ways.
The model we considered is given by
\begin{equation}
H_{S}=\frac{\omega}{2}\sigma_{z},
\end{equation}
\begin{equation}
L_{b}=\sigma_{z},\quad L_{f}=\sigma_{-}.
\end{equation}
According to the general discussion in subsection \ref{sub:IIIA},
the NMQSD equation for this model can be written as
\begin{equation}
\frac{\partial}{\partial t}|\psi_{t}\rangle=[-i\frac{\omega}{2}\sigma_{z}+\sigma_{-}\xi_{t}^{\ast}-\sigma_{+}\bar{Q}+\sigma_{z}z_{t}^{*}-\sigma_{z}\bar{O}]|\psi_{t}\rangle,
\end{equation}
In this example, the $O$ and $Q$ operators contains infinite order
of noise, therefore, for simplicity, we use the zeroth order functional expansion
to give the approximate zeroth order operators $O^{(0)}$ and $Q^{(0)}$. By assuming all 
the terms associated with noises are zero \cite{Xu2014}, the zeroth order $O^{(0)}$ and $Q^{(0)}$ operators 
can be obtained from Eqs.~(\ref{EqO}-\ref{EqQ}) as
\begin{equation}
O^{(0)}=\sigma_{z},
\end{equation}
\begin{equation}
Q^{(0)}=g(t,s)\sigma_{-},
\end{equation}
where the function $g(t,s)$ satisfies
\begin{equation}
\frac{\partial}{\partial t}g(t,s)=[i\omega+G(t)]g(t,s),
\end{equation}
where $G(t)=\int_{0}^{t}g(t,s)K_{f}(t,s)ds$. Finally, the corresponding
approximate master equation is
\begin{equation}
\frac{d}{dt}\rho=-i[\frac{\omega}{2}\sigma_{z},\rho]+\{G(t)[\sigma_{-}\rho,\sigma_{+}]+F(t)[\sigma_{z}\rho,\sigma_{z}]+\mathrm{H.c.}\}, \label{MEQ1q}
\end{equation}
where $F(t)=\int_{0}^{t}K_{b}(t,s)ds$. Different from the first example
where the model can be solved exactly, we show how to use the zeroth
order (for higher order expansion, see Ref. \cite{Xu2014}) approximation to derive an approximate master
equation in this second example. It is worth noting that Eq. (\ref{MEQ1q}) still contains incomplete non-Markovian information, although it is derived from the zeroth order 
approximation. In the Markov limit, the correlation functions $K_b(t,s)$ and $K_f(t,s)$ are all $\delta$-functions, and the coefficients are no longer 
time-dependent but reduced to constants. The zeroth-order approximation can partially capture the non-Markovian features as a way to improve
 the Markov approximation. Actually, in the real application
of the NMQSD approach, this systematic approximation method is shown to be very useful since the exact $O$ and $Q$ are often difficult to find
in many realistic models. With this approximation approach, one can still solve these non-Markovian problems with satisfactory accuracy.

\section{\label{sec:IV}Anti-commutative Case}

\subsection{\label{sub:IV-A}General Stochastic Schr\"{o}dinger Equation}

After discussing the commutative hybrid bath, we will consider the
case that the system is assumed to anti-commutes with the fermionic bath, which often
describes an electronic system such as a quantum dot system. Following a similar
procedure, we can also derive the NMQSD equation for the anti-commutative
hybrid bath as
\begin{eqnarray}
\frac{\partial}{\partial t}|\psi_{t}\rangle & = & -i\langle z^{*},\xi^{*}|H_{tot}^{int}(t)|\psi_{tot}(t)\rangle\nonumber \\
 & = & [-iH_{S}-L_{f}\xi_{t}^{\ast}-L_{f}^{\dagger}\int dsK_{f}(t,s)\frac{\delta_{l}}{\delta\xi_{s}^{\ast}}\nonumber \\
 &  & +L_{b}z_{t}^{\ast}-L_{b}^{\dagger}\int dsK_{b}(t,s)\frac{\delta}{\delta z_{s}^{\ast}}]\left\vert \psi_{t}\right\rangle .\label{QSDAC}
\end{eqnarray}
Since the operators $L_{f}$ typically anti-commutes with the fermionic
bath, the derivation is slightly different from the commutative case
\cite{ChenFB,ShiFB,ZhaoFB}. As a result, there is a minor difference
between Eq.~(\ref{QSDC}) and Eq.~(\ref{QSDAC}). The bosonic noise
$z_{t}^{*}$, the fermionic niose $\xi_{t}^{*}$ and the corresponding
correlation functions $K_{b}(t,s)$ and $K_{f}(t,s)$ are all defined
in the same way. Similarly, we can also define the bosonic $O$ operator
and the fermionic $Q$ operator, however, they satisfy different differential
equations in the anti-commutative case as 
\begin{eqnarray}
\frac{\partial}{\partial t}O & = & [-iH_{s}-L_{f}\xi_{t}^{\ast}-L_{f}^{\dagger}\bar{Q}+L_{b}z_{t}^{*}-L_{b}^{\dagger}\bar{O},O]\nonumber \\
 &  & -L_{b}^{\dagger}\frac{\delta}{\delta z_{s}^{*}}\bar{O}-L_{f}^{\dagger}\frac{\delta}{\delta z_{s}^{*}}\bar{Q},
\end{eqnarray}

\noindent 
\begin{eqnarray}
\frac{\partial}{\partial t}Q & = & [-iH_{s},Q]+[L_{f}\xi_{t}^{*},Q]+[L_{b}z_{t}^{*},Q]\nonumber \\
 &  & -L_{f}^{\dagger}\bar{Q}Q+QL_{f}^{\dagger}\bar{Q}-L_{b}^{\dagger}\bar{O}Q+QL_{f}^{\dagger}\bar{O}\nonumber \\
 &  & -L_{b}^{\dagger}\frac{\delta_{l}}{\delta\xi_{s}^{\ast}}\bar{O}+L_{f}^{\dagger}\frac{\delta_{l}}{\delta\xi_{s}^{*}}\bar{Q},
\end{eqnarray}

\noindent with the initial conditions 
\begin{equation}
O(t,t,z^{*},\xi^{*})=L_{b},
\end{equation}
\begin{equation}
Q(t,t,z^{*}\xi^{*})=L_{f}.
\end{equation}
Then, the density matrix can be also reproduced as
\begin{equation}
\rho(t)=\langle\langle P_{t}\rangle_{f}\rangle_{b},
\end{equation}
and the master equation can be derived as
\begin{eqnarray}
\frac{d}{dt}\rho & = & -i[H_{S},\rho]+[L_{b},\langle\langle P_{t}\bar{O}^{\dagger}\rangle_{f}\rangle_{b}]+[\langle\langle\bar{Q}P_{t}\rangle_{f}\rangle_{b},L_{b}^{\dagger}]\nonumber \\
 &  & +[L_{f},\langle\langle P_{t}\bar{Q}^{\dagger}(-\xi)\rangle_{f}\rangle_{b}]+[\langle\langle\bar{Q}P_{t}\rangle_{f}\rangle_{b},L_{f}^{\dagger}].\label{MEQAC}
\end{eqnarray}
In the derivation of the master equation, an anti-commutative version
of the Novikov theorem \cite{ShiFB,ChenFB} has been used. It is different
from either the commutative version of Novikov theorem for fermionic
bath \cite{ZhaoFB} or the one for bosonic bath \cite{Yu1999}. Similarly,
when $O$ and $Q$ are noise-independent, the master equation is reduced
to
\begin{equation}
\frac{d}{dt}\rho=-i[H_{S},\rho]+\{[\bar{Q}\rho,L_{f}^{\dagger}]+[\bar{O}\rho,L_{b}^{\dagger}]+\mathrm{H.c.}\}.
\end{equation}

\subsection{Example 3: Quantum dot in a hybrid bath}

\label{sec:IV-B}

In order to show the details of solving an anti-commutative hybrid
bath problem, we consider a specific example that is the Anderson
model in a bosonic environment (see Ref. \cite{Chung} for example).
In this particular example the general Hamiltonian Eq.~(\ref{Hybrid})
becomes 
\begin{equation}
H_{S}=\varepsilon d^{\dagger}d,
\end{equation}
describing the quantum dot,
\begin{equation}
H_{B}=\sum_{k,i=L,R}[\epsilon(k)-\mu_{i}]c_{ki}^{\dagger}c_{ki}+\sum_{r}\omega_{r}b_{r}^{\dagger}b_{r},
\end{equation}
 describing the two fermionic baths (``$L$'' and ``$R$'') and one
phonon bath, and 
\begin{equation}
H_{I}=\sum_{k,i=L,R}t_{k,i}c_{ki}^{\dagger}d+\mathrm{H.c.}+\sum_{r}\lambda_{r}(d^{\dagger}d-\frac{1}{2})(b_{r}+b_{r}^{\dagger}),\label{HI}
\end{equation}
describing the transport process between two fermionic bath and the dissipation process caused by a bosonic bath.

In the finite temperature case \cite{Yu-FiniteT}, we need to introduce
two fictitious bath ``$a_{L}$'' and ``$a_{R}$'' with the negative eigen-frequencies
as:
\begin{eqnarray}
H & = & H_{S}+\sum_{k,i=L,R}[\epsilon(k)-\mu_{i}]c_{ki}^{\dagger}c_{ki}+\{t_{ki}c_{ki}^{\dagger}d+ \mathrm{H.c.}\}\nonumber \\
 &  & +\sum_{r}\lambda_{r}(d^{\dagger}d-\frac{1}{2})(b_{r}+b_{r}^{\dagger})+\sum_{r}\Omega_{r}b_{r}^{\dagger}b_{r}\nonumber \\
 &  & +\sum_{k,i=L,R}-[\epsilon(k)-\mu_{i}]a_{ki}^{\dagger}a_{ki}.
\end{eqnarray}
Then, performing the Bogoliubov transformation
\begin{equation}
c_{ki}=\sqrt{1-\bar{n}_{ki}}c_{ki}^{\prime}+\sqrt{\bar{n}_{ki}}a_{ki}^{\prime\dagger}\quad(i=L,R),
\end{equation}
\begin{equation}
a_{ki}=\sqrt{1-\bar{n}_{ki}}a_{ki}^{\prime}-\sqrt{\bar{n}_{ki}}c_{ki}^{\prime\dagger}\quad(i=L,R),
\end{equation}
the Hamiltonian become
\begin{eqnarray}
H & = & H_{S}+\sum_{k,i=L,R}[\epsilon(k)-\mu_{i}]c_{ki}^{\prime\dagger}c_{ki}^{\prime}\nonumber \\
 &  & +\{t_{ki}(\sqrt{1-\bar{n}_{ki}}c_{ki}^{\prime\dagger}+\sqrt{\bar{n}_{ki}}a_{ki}^{\prime})d+ \mathrm{H.c.}\}\nonumber \\
 &  & +\sum_{r}\lambda_{r}(d^{\dagger}d-\frac{1}{2})(b_{r}+b_{r}^{\dagger})+\sum_{r}\Omega_{r}b_{r}^{\dagger}b_{r}\nonumber \\
 &  & +\sum_{k,i=L,R}-[\epsilon(k)-\mu_{i}]a_{ki}^{\prime\dagger}a_{ki}^{\prime}.
\end{eqnarray}
Redefining $\omega_{ki}=\epsilon(k)-\mu_{i}$, $g_{ki}=t_{ki}\sqrt{1-\bar{n}_{ki}},\; f_{ki}=t_{ki}\sqrt{\bar{n}_{ki}}$,
then, in the interaction picture, the Hamiltonian can be written as
\begin{align}
H_{int}(t) & =H_{S}+\sum_{r}\lambda_{r}(d^{\dagger}d-\frac{1}{2})(b_{r}e^{-i\Omega_{r}t}+b_{r}^{\dagger}e^{i\Omega_{r}t})\nonumber \\
 & +\{\sum_{k,i=L,R}g_{ki}e^{i\omega_{k}t}c_{ki}^{\prime\dagger}d+f_{ki}e^{i\omega_{k}t}a_{ki}^{\prime}d+ \mathrm{H.c.}\}.
\end{align}
By introducing one bosonic coherent state and two fermionic coherent
states as
\begin{equation}
|z\rangle=\prod_{r}\exp\{z_{r}b_{r}^{\dagger}\}|0\rangle,
\end{equation}
\begin{equation}
|\xi_{ia}\rangle=\prod_{k}(1-\xi_{kia}a_{ki}^{\prime\dagger})|0\rangle\;(i=L,R),
\end{equation}

\begin{equation}
|\xi_{ic}\rangle=\prod_{k}(1-\xi_{kic}c_{ki}^{\prime\dagger})|0\rangle\;(i=L,R).
\end{equation}
The stochastic state vector can be defined as
\begin{equation}
|\psi_{t}(z^{*},\xi_{La}^{*},\xi_{Ra}^{*},\xi_{Lc}^{*},\xi_{Rc}^{*})\rangle=\langle z^{*},\xi_{La}^{*},\xi_{Ra}^{*},\xi_{Lc}^{*},\xi_{Rc}^{*}|\psi_{tot}(t)\rangle.
\end{equation}
Following the general approach discussed in the last subsection, the
NMQSD equation for the stochastic state vector is derived as
\begin{eqnarray}
\frac{\partial}{\partial t}|\psi_{t}\rangle & = & H_{eff}|\psi_{t}\rangle,
\end{eqnarray}
where 
\begin{eqnarray}
H_{eff} & = & [-iH_{S}+d^{\dagger}\int_{0}^{t}dsK_{La}(t,s)\frac{\delta}{\delta\xi_{La,s}^{*}}+d\xi_{La,t}^{*}\nonumber \\
 &  & +d^{\dagger}\int_{0}^{t}dsK_{Ra}(t,s)\frac{\delta}{\delta\xi_{Ra,s}^{*}}+d\xi_{Ra,t}^{*}\nonumber \\
 &  & -d\int_{0}^{t}dsK_{Lc}(t,s)\frac{\delta}{\delta\xi_{Lc,s}^{*}}-d^{\dagger}\xi_{Lc,t}^{*}\nonumber \\
 &  & -d\int_{0}^{t}dsK_{Rc}(t,s)\frac{\delta}{\delta\xi_{Rc,s}^{*}}-d^{\dagger}\xi_{Rc,t}^{*}\nonumber \\
 &  & -b^{\dagger}\int_{0}^{t}ds\alpha(t,s)\frac{\delta}{\delta z_{s}^{*}}+dz_{t}^{*}].
\end{eqnarray}
In this equation, we introduced five noises as
\begin{equation}
z_{t}^{*}=-i\sum_{r}z_{r}^{*}e^{-i\Omega_{r}t},
\end{equation}
\begin{equation}
\xi_{ia,t}^{*}=-i\sum_{k}\xi_{kia}^{*}e^{-i\omega_{k}t},\;(i=L,R),
\end{equation}

\begin{equation}
\xi_{ic,t}^{*}=-i\sum_{k}\xi_{kic}^{*}e^{-i\omega_{k}t},\;(i=L,R),
\end{equation}
and the corresponding correlation functions are
\begin{equation}
\alpha(t,s)=\sum_{r}\lambda_{r}^{2}e^{-i\Omega_{r}(t-s)},
\end{equation}
\begin{equation}
K_{ia}(t,s)=\sum_{k}g_{ki}^{2}e^{-i\omega_{k}(t-s)}\quad(i=L,R),
\end{equation}
\begin{equation}
K_{ic}(t,s)=\sum_{k}f_{ki}^{2}e^{i\omega_{k}(t-s)}\quad(i=L,R),
\end{equation}
Among the noises above, $z_{t}^{*}$ is a complex Gaussian noise,
$\xi_{ia,t}^{*}$ and $\xi_{ic,t}^{*}$ are Grassmann Gaussian noises.
They satisfy the following statistical relations
\begin{equation}
\langle z_{t}\rangle_{b}=\langle z_{t}^{*}\rangle_{b}=0,
\end{equation}
\begin{equation}
\langle z_{t}^{*}z_{s}\rangle_{b}=\alpha(t,s),
\end{equation}
\begin{equation}
\langle\xi_{ia,t}^{*}\rangle_{f}=\langle\xi_{ia,t}\rangle_{f}=\langle\xi_{ic,t}^{*}\rangle_{f}=\langle\xi_{ic,t}\rangle_{f}=0,
\end{equation}
\begin{equation}
\langle\xi_{ia,t}^{*}\xi_{ia,s}\rangle_{f}=K_{ia}(t,s),\;\langle\xi_{ic,t}^{*}\xi_{ic,s}\rangle_{f}=K_{ic}(t,s).
\end{equation}

Following the technique discussed in subsection \ref{sub:IV-A}, the
time dependent operators $O$ and $Q$ are defined as
\begin{equation}
\frac{\delta}{\delta z_{s}^{*}}|\psi_{t}\rangle=O(t,s,z^{*})|\psi_{t}\rangle,
\end{equation}
\begin{equation}
\frac{\delta}{\delta\xi_{ia,s}^{*}}|\psi_{t}\rangle=Q_{ia}(t,s,\xi_{ia}^{*})|\psi_{t}\rangle\quad(i=L,R),
\end{equation}
\begin{equation}
\frac{\delta}{\delta\xi_{ic,s}^{*}}|\psi_{t}\rangle=Q_{ic}(t,s,\xi_{ic}^{*})|\psi_{t}\rangle\quad(i=L,R),
\end{equation}
and the the zeroth order approximation gives the solution of these
operators as
\begin{equation}
O\approx f_{1}(t,s)d^{\dagger}d,
\end{equation}
\begin{equation}
Q_{ic}\approx f_{ic}(t,s)d\;(i=L,R),
\end{equation}
\begin{equation}
Q_{ia}\approx f_{ia}(t,s)d^{\dagger}\;(i=L,R),
\end{equation}
while the coefficients satisfy
\begin{equation}
\frac{\partial}{\partial t}f_{1}(t,s)=0,
\end{equation}
\begin{equation}
\frac{\partial}{\partial t}f_{Lc}(t,s)=(i\varepsilon+F_{1}+F_{La}+F_{Ra}+F_{Lc}+F_{Rc})f_{Lc},
\end{equation}
\begin{equation}
\frac{\partial}{\partial t}f_{Rc}(t,s)=(i\varepsilon+F_{1}+F_{La}+F_{Ra}+F_{Lc}+F_{Rc})f_{Rc},
\end{equation}
\begin{equation}
\frac{\partial}{\partial t}f_{La}(t,s)=(-i\varepsilon-F_{1}-F_{La}-F_{Ra}-F_{Lc}-F_{Rc})f_{La},
\end{equation}
\begin{equation}
\frac{\partial}{\partial t}f_{Ra}(t,s)=(-i\varepsilon-F_{1}-F_{La}-F_{Ra}-F_{Lc}-F_{Rc})f_{Ra},
\end{equation}
where $F_{1}=\int_{0}^{t}\alpha(t,s)f_{1}(t,s)ds$, $F_{Lc}=\int_{0}^{t}K_{Lc}(t,s)f_{Lc}(t,s)ds$,
$F_{Rc}=\int_{0}^{t}K_{Rc}(t,s)f_{Rc}(t,s)ds$, $F_{La}=\int_{0}^{t}K_{La}(t,s)f_{La}(t,s)ds$,
$F_{Ra}=\int_{0}^{t}K_{Ra}(t,s)f_{Ra}(t,s)ds$. Finally, the master
equation is derived as
\begin{align}
\frac{\partial}{\partial t}\rho & =-i\varepsilon[d^{\dagger}d,\rho]+\{(F_{Lc}+F_{Rc})[d\rho,d^{\dagger}]\nonumber \\
 & +(F_{La}+F_{Ra})[d,d^{\dagger}\rho]+F_{1}[d^{\dagger}d\rho,d^{\dagger}d]+\mathrm{H.c.}\}.\label{MEQdot}
\end{align}
In the third example, the hybrid NMQSD approach is applied to a very important  system called 
Anderson model embedded in a bosonic dephasing environment.
First, we show how to map a finite temperature problem into a zero
temperature problem to apply the hybrid NMQSD approach in finite temperature
case. More important, we show the hybrid NMQSD approach provides us
a powerful tool to investigate the dynamics of a quantum system in
a non-Markovian regime. With the NMQSD approach, open quantum systems
coupled to a hybrid bath such as the example discussed above can be solved systematically
in non-Markovian regimes. Typically, in the Markov case, all the coefficients
in the master equation, $F_{Lc}$, $F_{Rc}$, $F_{La}$, $F_{Ra}$,
and $F_{1}$ are constants. However, in Eq.~(\ref{MEQdot}), those
coefficients are time-dependent, which reflects the non-Markovian
behavior even if the zeroth-order $O$ and $Q$ operators are employed. The higher-order non-Markovian approximations
can be implemented in a similar way.

\subsection{Fermionic Bath vs. Bosonic Bath}

The master equation derived in Eq.~(\ref{MEQdot}) can be used to
illustrate the difference between the fermionic bath and bosonic bath.
For this purpose, two parameters in the original Hamiltonian are specified to describe the coupling
strength to the fermionic bath and bosonic bath. As it is shown in
Eq.~(\ref{HI}), $t_{ki}$ determine how strong the interaction between
the system and fermionic bath, and $\lambda_{r}$ determine the strength
of the coupling to bosonic bath. In the numerical simulation, we will
introduce $c_{f}$ and $c_{b}$ to control the global coupling strengths
for fermionic bath and bosonic bath respectively. Namely, we replace
$t_{ki}$ by $\sqrt{c_{f}}t_{ki}$ and $\lambda_{r}$ by $\sqrt{c_{b}}\lambda_{r}$.
Then, these two parameters reflect the global coupling strength.
For example, if we take $c_{b}=0$, then the bosonic bath is switched 
off, and we can observe the evolution without presence of the bosonic
bath. In the numerical simulations, we use four Ornstein-Uhlenbeck
noises $K_{mn}(t,s)=\frac{\Gamma_{mn}}{2}\exp[(-\gamma_{mn}+i\phi_{mn})|t-s|]$
($m=L,R$; $n=a,c$) to model the correlation functions $K_{La}$,
$K_{Lc}$, $K_{Ra}$, $K_{Rc}$. The parameters are chosen as $\Gamma_{Lc}=0.017$,
$\gamma_{Lc}=0.3$, $\phi_{Lc}=1.1$, $\Gamma_{Rc}=0.034$, $\gamma_{Rc}=0.5$,
$\phi_{Rc}=1.65$, $\Gamma_{La}=0.012$, $\gamma_{La}=0.4$, $\phi_{La}=0.75$,
$\Gamma_{Ra}=0.044$, $\gamma_{Ra}=0.45$, $\phi_{Ra}=1.2$.

\noindent 
\begin{figure}
\begin{centering}
\includegraphics[width=0.95\columnwidth]{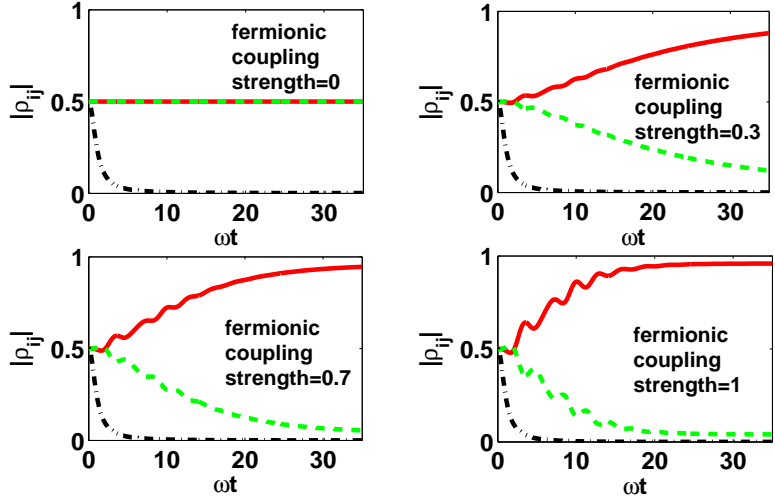}
\par\end{centering}

\caption{(Color online) Time evolution for different coupling strengths of fermionic
bath. The coupling strength of bosonic bath is fixed as 1. The red
(solid), green (dashed), and black (dash-dotted) curves are the elements
of density matrix $|\rho_{11}|$, $|\rho_{22}|$, $|\rho_{12}|$ respectively.}

\label{FB}
\end{figure}

\begin{figure}
\begin{centering}
\includegraphics[width=0.95\columnwidth]{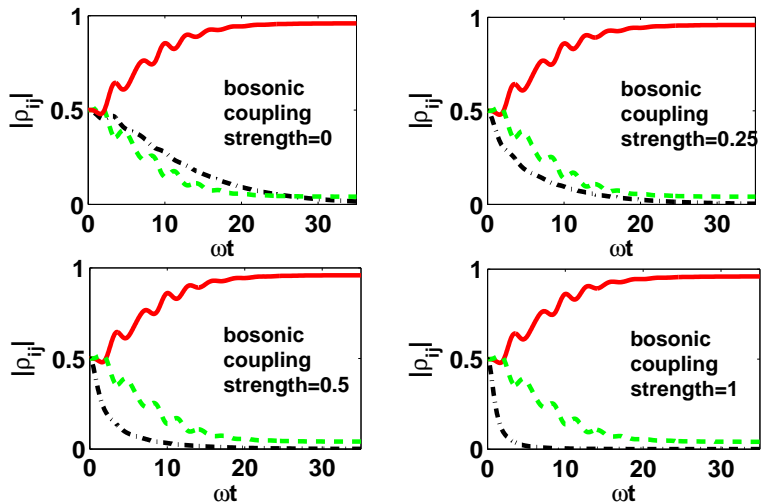}
\par\end{centering}

\caption{(Color online) Time evolution for different coupling strength of bosonic
bath. The coupling strength of fermionic bath is fixed as 1. The red
(solid), green (dashed), and black (dash-dotted) curves are the elements
of density matrix $|\rho_{11}|$, $|\rho_{22}|$, $|\rho_{12}|$ respectively.}

\label{BB}
\end{figure}

Fig.~\ref{FB} and \ref{BB} clearly show the different influences of
bosonic bath and fermionic bath on the system. Generally, For the model
under consideration, the fermionic bath
contributes to both of the energy dissipation and the decoherence, while the bosonic bath mainly
contributes to the dephasing process. From Fig.~\ref{FB}, we can
see that the dephasing process (off-diagonal elements)  remains almost the same
while changing the fermionic coupling strength. On the contrary, the
dissipative process is significantly modified. From Fig.~\ref{BB}, we
can see the dissipative process is not significantly affected by changing the bosonic
coupling strength, as a compassion, the dephasing rate is affected.
These results can be also predicted by analyzing the master equation
or Hamiltonian. Since the coupling form of the bosonic bath is a dephasing
type,  therefore it will fundamentally affect the dephasing process. In a similar fashion,
the coupling form of the fermionic bath is expected to affect the dissipative process.

\section{Conclusion}

\label{sec:V}

In this paper, we have developed a stochastic Schr\"odingier equation for
the open systems interacting with a hybrid environment containing both
bosons and fermions.  By combining the bosonic and fermionic NMQSD
approaches, two types of noises are used simultaneously to derive
the NMQSD equation for the hybrid bath case.  As a simple application, 
we show that the corresponding non-Markovian master equation for the hybrid bath case can be
recovered from the hybrid NMQSD equation. For more applications, two types of models are discussed 
including the commutative and anti-commutative cases.
In these examples, we have demonstrated  the consistency
between NMQSD approach and the ordinary quantum mechanics when the system and its environment are
relatively simple. Moreover, with these examples, 
we illustrate  the relationship between bosonic bath and fermionic bath, the exact and approximate
solutions of $O$ and $Q$ operators, and the different effects of
bosonic bath and fermionic bath. The hybrid NMQSD approach established
in this paper can be served as a convenient tool in the study of the dynamic 
evolution of the hybrid open
systems.  Particularly, it is helpful to investigate some early stage
evolution caused by memory effect of the environments since our approach
is systematically derived from the microscopic model which goes beyond
the standard Born-Markov approximation. 

\section*{Acknowledgements} We acknowledge grant support from the NSF
PHY-0925174, The NBRPC No.~2009CB929300,the NSFC Nos.~91121015 and the MOE No.~B06011.
J.Q.Y is supported by the NSAF (Grant Nos.~U1330201 and U1530401) and the National Basic Research 
Program of China (Grant Nos.~2016YFA0301201 and 2014CB921401).

\end{document}